\documentclass[a4paper, 12pt]{article}

\usepackage[sort&compress]{natbib}
\bibpunct{(}{)}{;}{a}{}{,} 

\usepackage{amsthm, amsmath, amssymb, mathrsfs, multirow, url, subfigure}
\usepackage{graphicx} 
\usepackage{ifthen} 
\usepackage{amsfonts}
\usepackage[usenames]{color}
\usepackage{fullpage}

\theoremstyle{plain}

\theoremstyle{definition}

\theoremstyle{remark}

\newtheorem*{astep}{A-step}
\newtheorem*{pstep}{P-step}
\newtheorem*{cstep}{C-step}

\newcommand{\prob}{\mathsf{P}}

\newcommand{\pl}{\mathsf{pl}}

\newcommand{\unif}{{\sf Unif}}

\newcommand{\YY}{\mathbb{Y}}
\newcommand{\UU}{\mathbb{U}}

\renewcommand{\S}{\mathcal{S}}

\title{Exact prior-free probabilistic inference in a class of non-regular models}
\author{Ryan Martin \\
Department of Statistics \\
North Carolina State University \\
\url{rgmarti3@ncsu.edu} \\
\mbox{} \\
Yi Lin \\
Department of Mathematics, Statistics, and Computer Science \\
University of Illinois at Chicago \\
\url{ylin46@uic.edu} 
}
\date{\today}

\begin{document}

\maketitle 

\begin{abstract}  
The use of standard statistical methods, such as maximum likelihood, is often justified based on their asymptotic properties.  For suitably regular models, this theory is standard but, when the model is non-regular, e.g., the support depends on the parameter, these asymptotic properties may be difficult to assess.  Recently, an inferential model (IM) framework has been developed that provides valid prior-free probabilistic inference without the need for asymptotic justification.  In this paper, we construct an IM for a class of highly non-regular models with parameter-dependent support.  This construction requires conditioning, which is facilitated through the solution of a particular differential equation.  We prove that the plausibility intervals derived from this IM are exact confidence intervals, and we demonstrate their efficiency in a simulation study.  

\smallskip

\end{abstract}

\section{Introduction}
\label{S:intro}  

For suitably regular problems, there is very little difference between the inferences obtained by various methods, at least for relatively large samples.  Indeed, under regularity conditions, the sampling distribution of the maximum likelihood estimator is asymptotically normal, and the Bayesian posterior takes the same asymptotically normal shape, regardless of the choice of prior.  The conclusions reached by using other non-standard approaches, such as generalized fiducial \citep[e.g.,][]{hannig2009, hannig.review.2015} and inferential models \citep{imbook}, would also be very similar to the likelihood and Bayesian conclusions in these regular problems.  However, outside the class of regular models, it is less clear that all the available methods will perform similarly.  For this reason, one could argue that perhaps the best examples to compare the quality of various approaches are non-regular models, particularly with small samples.  In this paper, we investigate some non-regular problems using the inferential model (IM) framework.  The initial work in this area focused mainly on the classical statistical inference problems, but some more recent work \citep[e.g.,][]{jin.li.jin.2015, gim} has investigated the IM approach in some non-standard problems.  The present paper continues this trend.   

A distinguishing feature of the IM approach is that it provides valid posterior probabilistic inference, in the form of a plausibility function, without the need to introduce artificial prior information.  The key idea is to put primary emphasis on a set of unobservable auxiliary variables which connect, through the model, to the observable data and parameter of interest.  The plausibility function to be used for inference obtains by predicting the unobserved auxiliary variables using a suitable random set.  Aside from providing meaningful probabilistic inference without priors, the IM's guaranteed validity property, Equation \eqref{eq:valid.pl}, implies that it also can be used to construct decision procedures with guaranteed frequentist error rate control.  In particular, a suitably constructed IM yields plausibilty intervals---the IM analogue of confidence intervals---whose coverage probability is guaranteed to be at least the nominal level, and this holds for all sample sizes and without regularity conditions.  But hitting the nominal level is only the first criterion, we also want the inference to be efficient.  It turns out that one of the keys to the IM's efficiency is the dimension of the auxiliary variable.  Reducing that dimension requires care in combining information coming from each data source.  For the non-regular models to be considered here, where sufficiency is not fully satisfactory, dimension reduction will be accomplished using a conditioning argument that involves solving a suitable differential equation.  We show that, for a highly non-regular model, the IM approach yields inference which is at least as efficient as the available default-prior Bayes and generalized fiducial methods.  The Bayes and fiducial solutions are very good, so the fact that the IM intervals are provably exact and no less efficient suggests that the IM solution is the best available for this problem.   

The remainder of the paper is organized as follows.  Section~\ref{S:review} gives a review of the IM approach with an illustration in a classical non-regular example, namely, $\unif(\theta, \theta+1)$.  In Section~\ref{S:nonregular}, we construct an IM for a class of non-regular models, i.e., $\unif(a(\theta), a(\theta) + b(\theta))$, for suitable functions $a$ and $b$, and we show that the corresponding plausibility intervals for $\theta$ have exactly the nominal coverage probability.  Numerical comparisons to a default-prior Bayes and a generalized fiducial approach are made, with the conclusion that the IM's provably exact results do not come at the cost of loss of efficiency.  Some concluding remarks are made in Section~\ref{S:discuss}.

\section{Review of IMs}
\label{S:review}

\subsection{IM construction and properties}
\label{SS:construction}

Let $Y \in \YY$ be the observable data and denote the sampling model by $\prob_{Y|\theta}$, indexed by a parameter $\theta \in \Theta$.  A critical component of the IM framework is an association between $Y$, $\theta$, and an unobservable auxiliary variable $U \in \UU$ with distribution $\prob_U$.  For this, we express the sampling model $Y \sim \prob_{Y|\theta}$ as follows: 
\begin{equation}
\label{eq:amodel}
Y = a(U,\theta), \quad \text{where} \quad U \sim \prob_U.  
\end{equation}
The key point behind this setup is that there exists a value $u^\star$ such that $y=a(u^\star, \theta)$, where $y$ is the observed value of $Y$ and $\theta$ is the true value of the parameter.  We consider the following basic three-step construction of an IM, which is based on the idea that inference on $\theta$, for given $Y=y$, can be achieved by shifting focus to the unobserved value $u^\star$ of the auxiliary variable $U$.  

\begin{astep}
Associate the observed data $Y=y$ and parameter $\theta$ for each value $u$ of the auxiliary variable $U$.  This results in a collection of subsets of $\Theta$:
\[ \Theta_y(u) = \{\theta: y = a(u,\theta)\}, \quad u \in \UU. \]
\end{astep}

\begin{pstep}
Predict $u^\star$ with a valid random set $\S$ having distribution $\prob_\S$. 
\end{pstep}

\begin{cstep}
Combine $\Theta_y(\cdot)$ with $\S$ to get
\[ \textstyle \Theta_y(\S) = \bigcup_{u \in \S} \Theta_y(u). \]
The intuition behind this combination step is that $\Theta_y(\S)$ contains the true $\theta$ if and only if $\S$ contains the true $u^\star$; we can construct $\S$ so that the latter event has high $\prob_\S$-probability (see below), so the former event also has high $\prob_\S$-probability.  Then, for any assertion $A \subseteq \Theta$, summarize the evidence in the observed data $Y=y$ supporting the truthfulness of $A$ with the plausibility function, 
\begin{equation}
\label{eq:bel} 
\pl_y(A) = \prob_\S\{\Theta_y(\S) \cap A \neq \varnothing\}. 
\end{equation}
\end{cstep}

The above construction appears rather simple, but there are two hidden subtleties: one concerning the association in the A-step and the other concerning the choice of predictive random set $\S$ in the P-step. 
\begin{itemize}
\item Intuitively, between a a low- and high-dimensional auxiliary variable, the former is preferred.  There are two reasons for this, the main one being that the IM is most efficient when the dimension of the auxiliary variable equals that of the parameter.  In regular models, an argument based on sufficient statistics is enough to get the best possible reduction.  However, in the non-regular models of interest here, sufficiency alone will not be fully satisfactory.  We will employ a differential equation technique to help further reduce the dimension via conditioning; see Sections~\ref{SS:location} and \ref{S:nonregular}.  
\vspace{-2mm}
\item A general discussion of predictive random sets can be found in \citet{imbook} but, in this paper, we will need only some relatively simple and concrete ideas.  We will require that $\S \sim \prob_\S$ be \emph{valid} in the sense that 
\begin{equation}
\label{eq:prs.valid}
f_\S(U) \sim \unif(0,1) \quad \text{when $U \sim \prob_U$}, 
\end{equation}
where $f_\S(u) = \prob_\S(\S \ni u)$.  It turns out that this condition is easy to arrange in our case here, in part because the dimension-reduction steps outlined above will result in us having only a scalar auxiliary variable $U$ to contend with.  
\end{itemize}

To summarize, an IM starts with an association and a valid predictive random set $\S$, and its output is a plausibility function $\pl_y$ for inference about $\theta$.  \citet{imbasics} show that, if \eqref{eq:prs.valid} holds and $\Theta_y(\S)$ is non-empty with $\prob_\S$-probability~1 for all $y$, then 
\begin{equation}
\label{eq:valid.pl}
\sup_{\theta \in A} \prob_{Y|\theta}\{\pl_Y(A) \leq \alpha\} \leq \alpha, \quad \forall \; \alpha \in (0,1), \quad \forall \; A \subseteq \Theta. 
\end{equation}
This is a very general result, which says that the plausibility function is calibrated so that the corresponding inference is scientifically meaningful.  An important consequence of this validity property is that the $100(1-\alpha)$\% plausibility region 
\begin{equation}
\label{eq:pl.region}
\{\theta: \pl_y(\theta) > \alpha\}, 
\end{equation}
with $\pl_y(\theta) = \pl_y(\{\theta\})$, is a $100(1-\alpha)$\% confidence region.  Note that this conclusion is not based on asymptotics, nor does it require the model to be ``regular'' in any way.

\subsection{Illustration}
\label{SS:location}

To illustrate the general IM machinery, here we will work through a classical non-regular example.  Suppose that the observable data $Y=(Y_1,\ldots,Y_n)$ are iid $\unif(\theta, \theta+1)$; what follows can be easily modified for other suitably non-regular location parameter models.  Start by reducing to a minimal sufficient statistic, $X=(X_1,X_2)$, where $X_1 = \min(Y_i)$ and $X_2 = \max(Y_i)$.  Then we can write down a baseline association 
\begin{equation}
\label{eq:baseline.toy1}
X_i = \theta + U_i, \quad i=1,2, 
\end{equation}
where the unobservable auxiliary variable $U=(U_1,U_2)$ has the (joint) distribution of the minimum and maximum of $n$ iid $\unif(0,1)$ random variables.  The reduction provided by sufficiency here is not fully satisfactory since we prefer a one-dimensional auxiliary variable for inference on a one-dimensional parameter.  In the classical setting, it would be natural in such cases to condition on some feature of $X$, namely, an ancillary statistic.  Here, we provide a different take on this conditioning approach from an IM perspective. 

The jumping off point is the observation that there are two equations in \eqref{eq:baseline.toy1}, with only one parameter, so some feature of $U=(U_1,U_2)$ is actually observed and, therefore, to improve efficiency, we should condition on it.  It is easy enough to guess what that feature should be but, for the sake of illustration, we employ a differential equation-based technique described in \citet{imcond}.  Let $u_{x,\theta}=(u_{x,\theta}^1, u_{x,\theta}^2)$ be a solution to the system \eqref{eq:baseline.toy1} for a given $x=(x_1,x_2)$ and $\theta$.  Then a feature $\eta(u_{x,\theta})$ of $u_{x,\theta}$ is observed if it is not sensitive to changes in $\theta$, i.e., if 
\[ \frac{\partial \eta(u_{x,\theta})}{\partial \theta} = 0. \]
An easy chain-rule calculation in this case reveals that a solution to this differential equation is the difference, $\eta(u) = u_2 - u_1$.  Also, if we take $\tau(u) = u_1$, then it is easy to see that $u \mapsto (\tau(u), \eta(u))$ is one-to-one.  Then we can rewrite \eqref{eq:baseline.toy1} as 
\begin{equation}
\label{eq:baseline.toy2}
T(X) = \tau(U) + \theta \quad \text{and} \quad H(X) = \eta(U), 
\end{equation}
where $T(x) = x_1$ and $H(x) = x_2-x_1$; clearly, $x \mapsto (T(x), H(x))$ is one-to-one as well.  Note that, by construction, the second equation does not involve $\theta$, so it is natural to condition on this event.  In the IM context, this conditioning approach boils down to working with the conditional distribution of $\tau(U)$, given $\eta(U) = h$, where $h$ is the observed value of $H(X)$.  It is a routine exercise to show that 
\[ \tau(U) \mid \{\eta(U) = h\} \sim \unif(0, 1-h), \quad h \in (0,1). \]
From this, we can define a conditional association as 
\begin{equation}
\label{eq:cond.assoc.toy}
T(X) = \theta + (1-h) \, W, \quad W \sim \unif(0,1), \quad h=H(X). 
\end{equation}
This completes the A-step in the IM construction.  

For the P-step, we consider the the so-called ``default'' predictive random set, $\S$, for predicting unobserved auxiliary variables from $\unif(0,1)$, i.e., 
\begin{equation}
\label{eq:default}
\S = \{w: |w-\tfrac12| \leq |W-\tfrac12|\}, \quad W \sim \unif(0,1). 
\end{equation}
This choice is theoretically valid in the sense that \eqref{eq:prs.valid} holds \citep[e.g.,][Sec.~3]{imbasics} and leads to computationally simple IM output; it was also shown to be optimal in a certain sense for symmetric regular distributions, and similar properties are to be expected here in this symmetric non-regular case, but see Section~\ref{SS:pstep}.  

Finally, for the C-step, we get the new random set 
\begin{align*}
\Theta_x(\S \mid h) & = \bigl\{ \theta: \tfrac{1}{1-h}(T(x) - \theta) \in \S \bigr\} \\
& = \bigl\{\theta: | \tfrac{1}{1-h}(T(x) - \theta) - \tfrac12| \leq |W-\tfrac12| \bigr\}, \quad W \sim \unif(0,1). 
\end{align*}
From this point, any plausibility function calculation can be obtained by computing a relevant probability with respect to $W \sim \unif(0,1)$.  For example, the point-wise (conditional) plausibility function equals 
\begin{align*}
\pl_x(\theta \mid h) & = \prob_\S\{\Theta_x(\S \mid h) \ni \theta\} \\
& = \prob_W\bigl\{ |W-\tfrac12| \geq |\tfrac{1}{1-h}(T(x) - \theta) - \tfrac12| \bigr\} \\
& = 1 - | \tfrac{2}{1-h}(T(x) - \theta) - 1|. 
\end{align*}
From here, it is simple to read off the formula for the corresponding $100(1-\alpha)$\% (conditional) plausibility interval for $\theta$:
\[ \{\theta: \pl_x(\theta \mid h) \geq \alpha\} = \bigl[ T(x) - (1-h)(1-\tfrac{\alpha}{2}), \, T(x) - (1-h) \tfrac{\alpha}{2} \bigr]. \]
One can check directly that this $100(1-\alpha)$\% plausibility interval has (conditional and unconditional) frequentist coverage probability exactly equal to $1-\alpha$.  This is a consequence of the general validity properties of the (conditional) IM.     

For comparison, consider a default-prior Bayes approach where, as is customary for location parameters, we take a flat prior for $\theta$.  It is easy to check that the corresponding Bayesian posterior distribution for $\theta$ is 
\[ \theta \mid X \sim \unif(X_2 - 1, X_1). \]
Routine calculation shows that the corresponding equal-tailed $100(1-\alpha)$\% credible interval is exactly the plausibility interval above, again with $h$ equal to the observed value of $X_2-X_1$.  So, the Bayesian credible interval is also exact in a frequentist sense; a potential advantage, however, of the IM approach is that there is no need to introduce artificial prior information in order to get valid posterior inference.

\section{IMs for a class of non-regular models}
\label{S:nonregular}

\subsection{IM construction: A-step}

Consider an iid sample $Y=(Y_1,\ldots,Y_n)$ from a $\unif(a(\theta), a(\theta) + b(\theta))$ distribution, where the functions $a(\theta)$ and $b(\theta) > 0$ are known and sufficiently smooth, and inference on the unknown scalar parameter $\theta$ is required.  A sufficiency argument, as in the previous illustration, reduces the problem to a minimal sufficient statistic, $X=(X_1, X_2)$, the sample minimum and maximum, respectively.  This leads to a baseline association 
\begin{equation}
\label{eq:baseline}
X_i = b(\theta) \, U_i + a(\theta), \quad i=1,2, 
\end{equation}
where $U=(U_1,U_2)$ has the (joint) distribution of the sample minimum and maximum of $n$ iid $\unif(0,1)$ random variables.  As before, sufficiency does not provide a satisfactory reduction of the dimension of the auxiliary variable.

The first step towards reducing the dimension is to identify two pairs of one-to-one functions functions, $x \mapsto (T(x), H(x))$ and $u \mapsto (\tau(u), \eta(u))$, such that 
\[ T(X) = G\bigl( \tau(U), \theta \bigr) \quad \text{and} \quad H(X) = \eta(U), \]
where $G(\cdot, \theta)$ is a known function; as we will see below, we may need to allow $\eta=\eta_{\theta_0}$ and $H=H_{\theta_0}$ to depend on a particular value $\theta_0$ of $\theta$, a localization point.  To find $\eta$ (and $H$), we consider the differential equation
\begin{equation}
\label{eq:pde}
\frac{\partial \eta(u_{x,\theta})}{\partial \theta} \equiv \Bigl( \frac{\partial \eta(u)}{\partial u} \Bigr|_{u=u_{x,\theta}} \Bigr)^\top \, \frac{\partial u_{x,\theta}}{\partial \theta} = 0, 
\end{equation}
where $u_{x,\theta}$ is the solution to the system defined by \eqref{eq:baseline}, for fixed $(x,\theta)$, which, in this case, is given by 
\[ u_{x,\theta} = (u_{x,\theta}^1, u_{x,\theta}^2)^\top =  \Bigl( \frac{x_1-a(\theta)}{b(\theta)}, \frac{x_2-a(\theta)}{b(\theta)} \Bigr)^\top. \]
The derivative is 
\[ \frac{\partial u_{x,\theta}^i}{\partial \theta} = -\frac{a'(\theta)}{b(\theta)} - \frac{b'(\theta)}{b(\theta)} \, u_{x,\theta}^i, \quad i=1,2, \]
and it is easy to check that 
\[ \eta(u) = \eta_{\theta_0}(u) = \log\Bigl\{ \frac{b'(\theta_0) \, u_1 + a'(\theta_0)}{b'(\theta_0) \, u_2 + a'(\theta_0)} \Bigr\} \]
solves the differential equation \eqref{eq:pde}, when $\theta=\theta_0$.  Moreover, note that the baseline association is ``separable,'' i.e., the solution $\eta$ will not depend on a particular localization point $\theta_0$, if and only if the endpoint functions $a(\theta)$ and $b(\theta)$ are linear in $\theta$; a special case of this is the $\unif(\theta,\theta+1)$ example in Section~\ref{SS:location}.  Having identified $\eta=\eta_{\theta_0}$, the function $H=H_{\theta_0}$ is also determined: 
\begin{align*}
H(x) & = H_{\theta_0}(x) = \eta_{\theta_0}(u_{x,\theta_0}) = \log\Bigl\{ \frac{b'(\theta_0) \, u_{x,\theta_0}^1 + a'(\theta_0)}{b'(\theta_0) \, u_{x,\theta_0}^2 + a'(\theta_0)} \Bigr\} 
\end{align*}

It remains to specify a function $\tau$ such that $u \mapsto (\tau(u),\eta(u))$ is one-to-one.  Here we consider $\tau(u)=u_1$, and it is easy to check that, with this choice, the above map is one-to-one.  Indeed, if $v_1 = \tau(u)$ and $v_2 = \eta(u)$, then the inverse mapping is 
\[ u_1 = v_1 \quad \text{and} \quad u_2 = \frac{a'(\theta_0)}{b'(\theta_0)} (e^{-v_2}-1) + e^{-v_2} v_1. \]
Therefore, we can rewrite the baseline association \eqref{eq:baseline} as 
\[ X_1 = b(\theta) U_1 + a(\theta) \quad \text{and} \quad H_{\theta_0}(X) = \eta_{\theta_0}(U). \]
Since the second equation does not involve $\theta$, the general theory in Martin and Liu (2015) suggests to condition on that event.  This leads to a conditional association
\begin{equation}
\label{eq:conditional.association}
X_1 = b(\theta) F_{h_{\theta_0}}^{-1}(W) + a(\theta), \quad W \sim \unif(0,1), 
\end{equation}
where $F_{h}$ is the conditional distribution function of $V_1=\tau(U)$ given $V_2=\eta_{\theta_0}(U)$ takes value $h$, and $h_{\theta_0}$ is the observed value of $H_{\theta_0}(X)$.  Finally, it is a routine calculus exercise to show that 
\[ F_{h}(v_1) = \frac{\{a'(\theta_0) + b'(\theta_0) v_1\}^n - a'(\theta_0)^n}{[\{a'(\theta_0) + b'(\theta_0)\} e^h]^n - a'(\theta_0)^n}, \quad 0 \leq v_1 \leq \Bigl\{1 + \frac{a'(\theta_0)}{b'(\theta_0)} \Bigr\}e^h - \frac{a'(\theta_0)}{b'(\theta_0)}. \]
Note that this distribution has a density function which is strictly increasing on its support, similar to a ${\sf Beta}(a,1)$ density function, $a > 1$.

\subsection{IM construction: P- and C-steps}
\label{SS:pstep}

Having written down an association involving only a scalar auxiliary variable, $W$, in \eqref{eq:conditional.association}, we are now ready for the P-step.  It is tempting, as in Section~\ref{SS:location}, to take the default predictive random set in \eqref{eq:default}.  However, those optimality properties enjoyed by the default predictive random set hold only for (nearly) symmetric distributions; in this case, as mentioned at the end of the previous section, the (conditional) distribution in question is very skewed, so the default is not appropriate here.  Ideally, we want to choose a random set whose support matches up with the level sets of the auxiliary variable distribution.  In this case, this can be accomplished by taking a predictive random set for $W$ in \eqref{eq:conditional.association} as 
\[ \S = [W, 1], \quad W \sim \unif(0,1). \]
For the given conditional association and this predictive random set, the C-step yields 
\begin{align*}
\Theta_x(\S \mid h_{\theta_0}) & = \Bigl\{\theta: F_{h_{\theta_0}}\Bigl( \frac{x_1 - a(\theta)}{b(\theta)} \Bigr) \in \S \Bigr\} \\
& = \Bigl\{\theta: F_{h_{\theta_0}}\Bigl( \frac{x_1 - a(\theta)}{b(\theta)} \Bigr) \geq W \Bigr\}, \quad W \sim \unif(0,1). 
\end{align*}
From here, just like before, we can easily write down a pointwise (conditional) plausibility function 
\[ \pl_x(\theta \mid h_{\theta_0}) = F_{h_{\theta_0}}\Bigl( \frac{x_1 - a(\theta)}{b(\theta)} \Bigr). \]
We have an explicit formula for the distribution function $F_h$, so this can be directly calculated---no Monte Carlo methods are needed.  Inverting this distribution function for the purpose of obtaining plausibility intervals for $\theta$ can also be done explicitly.  

A question to be addressed here concerns the choice of the localization point $\theta_0$.  As discussed in \citet{imcond}, at least for the pointwise plausibility function and the corresponding plausibility intervals, there is no reason that we cannot take $\theta_0$ equal to the particular argument $\theta$.  That is, the plausibility function would be 
\[ \pl_x(\theta \mid h_\theta) = F_{h_{\theta}}\Bigl( \frac{x_1 - a(\theta)}{b(\theta)} \Bigr). \]
Plots of this function can easily be generated, to summarize uncertainty, and the corresponding $100(1-\alpha)$\% plausibility interval, 
\[ \{\theta: \pl_x(\theta \mid h_\theta) \geq \alpha\}, \]
has exactly the nominal coverage probability.  To prove this latter claim, \eqref{eq:conditional.association} implies that the conditional distribution of $\pl_X(\theta \mid H_\theta(X))$, given $H_\theta(X)$, is $\unif(0,1)$.  Therefore, 
\[ \prob_{X|\theta}\bigl\{ \pl_X(\theta \mid H_\theta(X)) \geq \alpha \mid H_\theta(X) \bigr\} = 1-\alpha. \]
Taking expectation with respect to the marginal distribution of $H_\theta(X)$ reveals that, as claimed, the unconditional coverage probability is also $1-\alpha$.

\subsection{Numerical examples}
\label{S:numerical}

A common example in the literature, of this general form, is $\unif(\theta, \theta^2)$, for $\theta > 1$.  This corresponds to $a(\theta) = \theta$ and $b(\theta) = \theta^2 - \theta$.  Bayesian approaches based on default priors are available for this problem, e.g., \citet[][Example~9]{bergerbernardosun2009} and \citet[Example~3]{shemyakin2014}.  Unlike for the regular scalar-parameter cases where Jeffreys prior is the agreed-upon default prior, for this non-regular problem, there are several priors that claim to be ``reference'' and no general non-asymptotic guarantees are available on the coverage probability of the corresponding credible intervals.  As an alternative to the default-prior Bayes approach, a generalized fiducial solution was recently worked out in \citet{hannig.review.2015}, which looks similar to a Bayesian solution, but with a data-dependent prior.  Again, no non-asymptotic properties of the corresponding fiducial confidence intervals are available.  The proposed conditional IM approach, on the other hand, provides plausibility regions which are provably exact.  Here we carry out some numerical experiments to compare the performance of the various methods.   

\begin{figure}
\begin{center}
\subfigure[IM plausibility function]{\scalebox{0.6}{\includegraphics{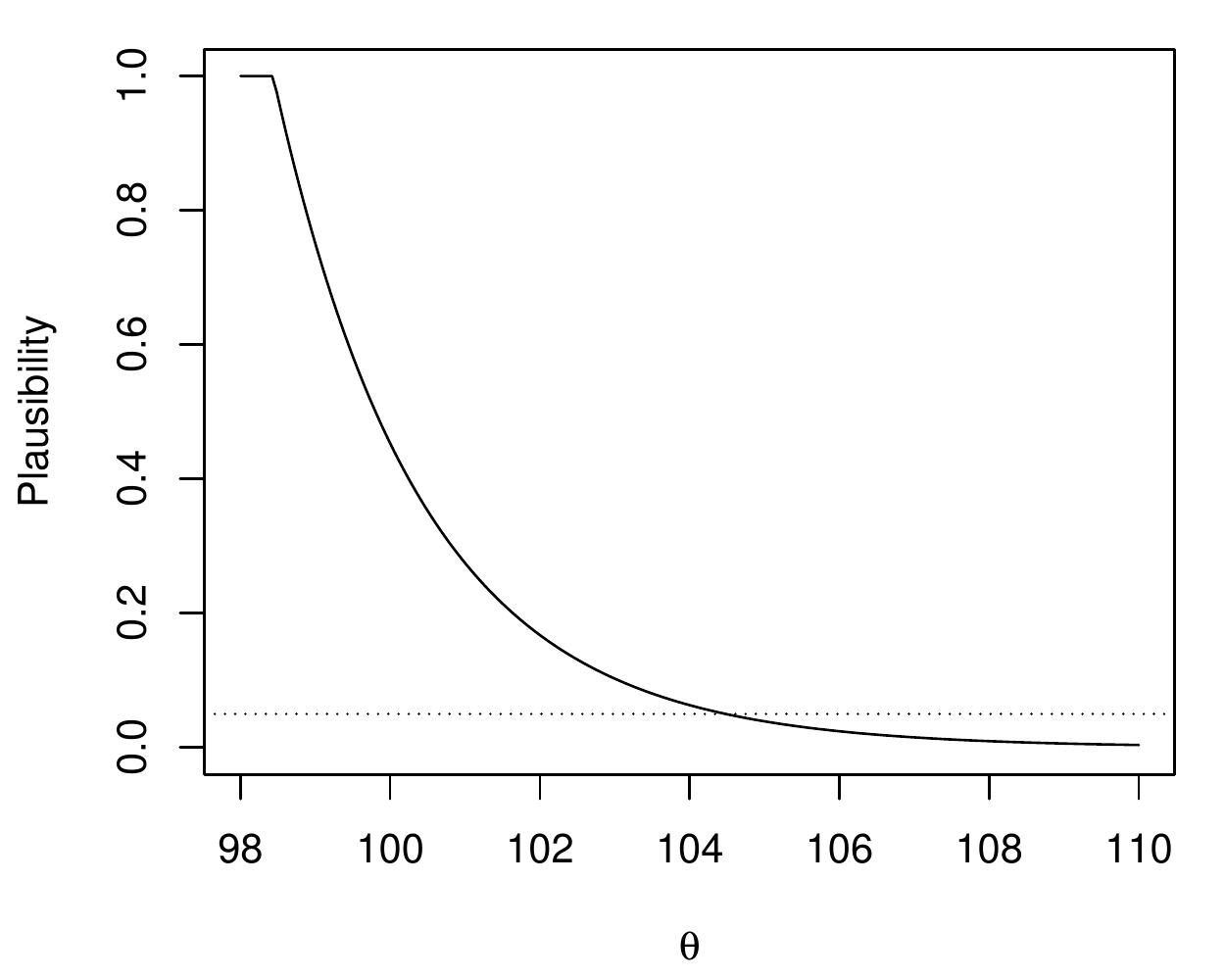}}}
\subfigure[Bayes/fiducial posterior density]{\scalebox{0.6}{\includegraphics{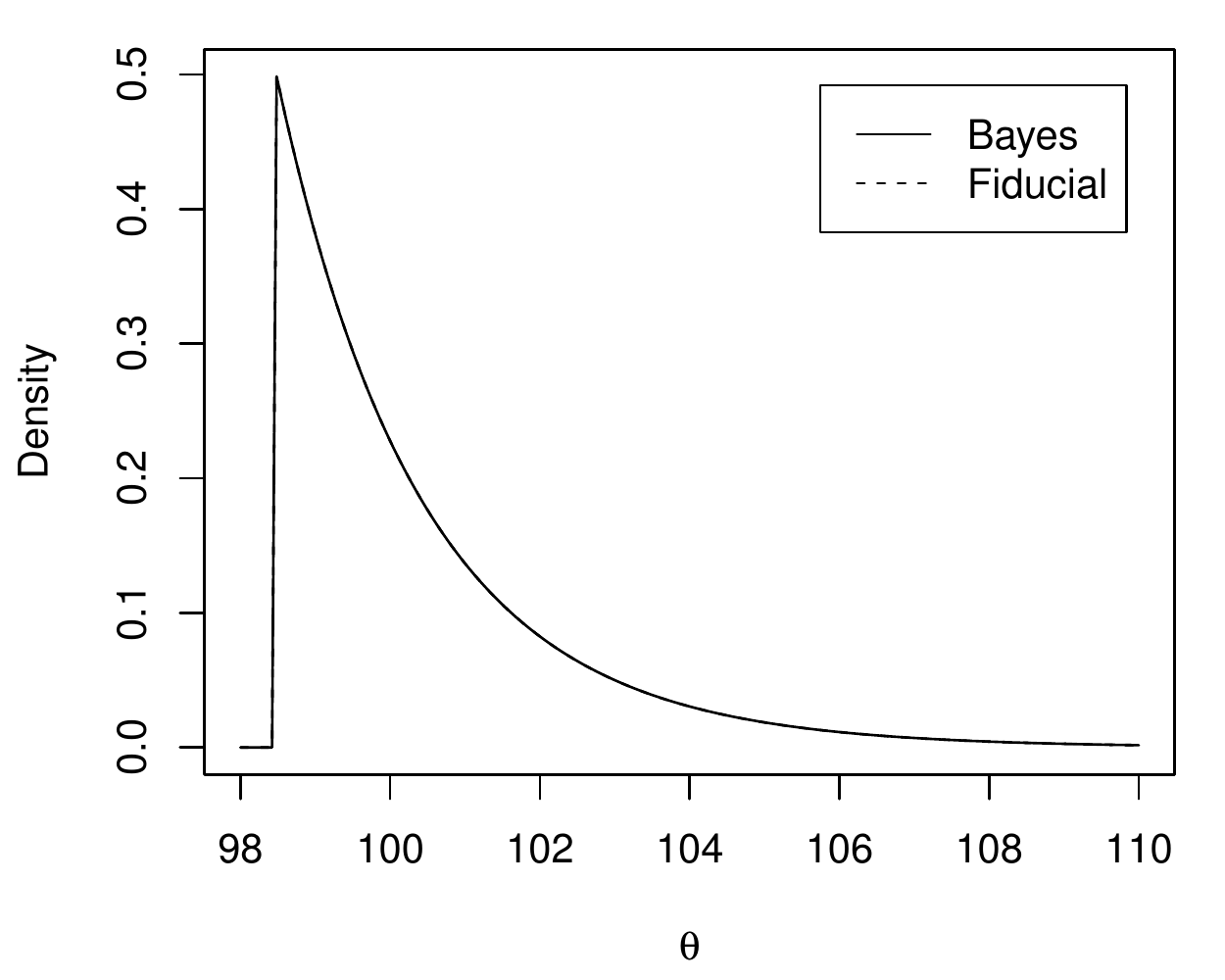}}}
\end{center}
\caption{Results for the $\unif(\theta, \theta^2)$ example based on data/sufficient statistics $X=(281.1, 9689.7)$.  The true value of $\theta$ used to generate these data was $\theta=100$.}
\label{fig:hannig}
\end{figure}

As a first illustration, we reconsider the example presented in \citet{hannig.review.2015}, where they simulate $n=25$ observations from $\unif(\theta, \theta^2)$, where $\theta=100$ is the true value.  They reported $X_1=281.1$ and $X_2=9689.7$ for the sample minimum and maximum, respectively.  Figure~\ref{fig:hannig}(a) plots $\pl_x(\theta \mid h_\theta)$ for these data; the horizontal line at $\alpha=0.05$ defines the 95\% plausibility region which, in this case, is $(98.44, 104.48)$.  This interval is effectively the same as the 95\% highest posterior density intervals based on the fiducial and default-prior \citep[cf.,][]{shemyakin2014} Bayesian posteriors plotted in Figure~\ref{fig:hannig}(b), which are indistinguishable.  The two-sided 95\% fiducial interval, $(98.49, 105.92)$, reported in \citet{hannig.review.2015}, agrees with the IM plausibility interval obtained by using the ``default''  random set \eqref{eq:default} in the P-step, which we argued was not appropriate.

To further compare the IM results with that of the default-prior Bayes and fiducial approaches, we carry out a simulation study.  We consider samples of size $n=2, 5, 10, 25$ from a $\unif(\theta, \theta^2)$ distribution, for $\theta=2, 5, 10, 25, 50, 100$.  Table~\ref{fig:coverage} shows the estimated coverage probability for the 95\% Bayes, fiducial, and intervals, based on 8000 Monte Carlo samples; and Table~\ref{fig:length} shows the estimated mean length of these two intervals.  In terms of coverage probability, all three intervals are very similar and within an acceptable range of the target 0.95.  In terms of length, the Bayesian intervals tend to be longer than the others, but the average lengths of the fiducial and IM intervals are quite similar.  Given that the IM-based intervals are provably exact and appear to be at least as efficient as the others in terms of length, we conclude that the IM approach is the better of the three for this problem.

\begin{table}
\begin{center}
\begin{tabular}{cccccccc}
\hline
& & \multicolumn{6}{c}{$\theta$} \\
\cline{3-8} 
Method & $n$ & 2 & 5 & 10 & 25 & 50 & 100 \\
\hline
Bayes & 2 & 0.954 & 0.951 & 0.955 & 0.953 & 0.952 & 0.955 \\
& 5 & 0.950 & 0.949 & 0.947 & 0.950 & 0.949 & 0.952 \\
& 10 & 0.951 & 0.953 & 0.949 & 0.949 & 0.952 & 0.953 \\
& 25 & 0.952 & 0.948 & 0.952 & 0.944 & 0.950 & 0.956 \\
\hline
Fiducial & 2 & 0.953 & 0.946 & 0.951 & 0.950 & 0.950 & 0.954 \\
& 5 & 0.950 & 0.948 & 0.946 & 0.950 & 0.949 & 0.951 \\
& 10 & 0.951 & 0.952 & 0.948 & 0.949 & 0.951 & 0.953 \\
& 25 & 0.952 & 0.948 & 0.952 & 0.944 & 0.950 & 0.956 \\
\hline
IM & 2 & 0.952 & 0.945 & 0.951 & 0.951 & 0.951 & 0.954 \\
& 5 & 0.949 & 0.948 & 0.946 & 0.950 & 0.949 & 0.951 \\
& 10 & 0.950 & 0.952 & 0.948 & 0.949 & 0.952 & 0.952 \\
& 25 & 0.952 & 0.948 & 0.952 & 0.944 & 0.950 & 0.955 \\
\hline
\end{tabular}
\end{center}
\caption{Estimated coverage probabilities for the 95\% Bayes, fiducial, and IM intervals in the $\unif(\theta,\theta^2)$ simulation study.  Estimates are based on 8000 Monte Carlo samples.}
\label{fig:coverage}
\end{table}

\begin{table}
\begin{center}
\begin{tabular}{cccccccc}
\hline
& & \multicolumn{6}{c}{$\theta$} \\
\cline{3-8} 
Method & $n$ & 2 & 5 & 10 & 25 & 50 & 100 \\
\hline 
Bayes & 2 & 0.6260 & 3.170 & 7.610 & 21.00 & 43.40 & 88.70 \\
& 5 & 0.2690 & 1.190 & 2.770 & 7.54 & 15.50 & 31.60 \\
& 10 & 0.1360 & 0.588 & 1.350 & 3.67 & 7.52 & 15.30 \\
& 25 & 0.0552 & 0.234 & 0.535 & 1.44 & 2.96 & 6.03 \\
\hline
Fiducial & 2 & 0.6170 & 3.080 & 7.420 & 20.70 & 43.00 & 88.30 \\
& 5 & 0.2680 & 1.180 & 2.760 & 7.51 & 15.40 & 31.60 \\
& 10 & 0.1360 & 0.586 & 1.350 & 3.66 & 7.51 & 15.30 \\
& 25 & 0.0552 & 0.234 & 0.535 & 1.44 & 2.96 & 6.02 \\
\hline
IM & 2 & 0.6130 & 3.070 & 7.430 & 20.80 & 43.10 & 88.00 \\
& 5 & 0.2670 & 1.180 & 2.770 & 7.53 & 15.40 & 31.40 \\
& 10 & 0.1360 & 0.587 & 1.350 & 3.66 & 7.51 & 15.20 \\
& 25 & 0.0552 & 0.234 & 0.535 & 1.44 & 2.96 & 5.98 \\
\hline
\end{tabular}
\end{center}
\caption{Estimated mean lengths of the 95\% Bayes, fiducial, and IM intervals in the $\unif(\theta,\theta^2)$ simulation study.  Estimates are based on 8000 Monte Carlo samples.}
\label{fig:length}
\end{table}

\ifthenelse{1=1}{}{

8000 Monte Carlo samples
n = 2, 5, 10, 25 (rows)
theta = 2, 5, 10, 25, 50, 100 (columns)

> print(signif(im.cvg, 3))
      [,1]  [,2]  [,3]  [,4]  [,5]  [,6]
[1,] 0.952 0.945 0.951 0.951 0.951 0.954
[2,] 0.949 0.948 0.946 0.950 0.949 0.951
[3,] 0.950 0.952 0.948 0.949 0.952 0.952
[4,] 0.952 0.948 0.952 0.944 0.950 0.955

>     print(signif(im.len, 3))
       [,1]  [,2]  [,3]  [,4]  [,5]  [,6]
[1,] 0.6130 3.070 7.430 20.80 43.10 88.00
[2,] 0.2670 1.180 2.770  7.53 15.40 31.40
[3,] 0.1360 0.587 1.350  3.66  7.51 15.20
[4,] 0.0552 0.234 0.535  1.44  2.96  5.98

>     print(signif(ba.cvg, 3))
      [,1]  [,2]  [,3]  [,4]  [,5]  [,6]
[1,] 0.954 0.951 0.955 0.953 0.952 0.955
[2,] 0.950 0.949 0.947 0.950 0.949 0.952
[3,] 0.951 0.953 0.949 0.949 0.952 0.953
[4,] 0.952 0.948 0.952 0.944 0.950 0.956

>     print(signif(ba.len, 3))
       [,1]  [,2]  [,3]  [,4]  [,5]  [,6]
[1,] 0.6260 3.170 7.610 21.00 43.40 88.70
[2,] 0.2690 1.190 2.770  7.54 15.50 31.60
[3,] 0.1360 0.588 1.350  3.67  7.52 15.30
[4,] 0.0552 0.234 0.535  1.44  2.96  6.03

>     print(signif(fi.cvg, 3))
      [,1]  [,2]  [,3]  [,4]  [,5]  [,6]
[1,] 0.953 0.946 0.951 0.950 0.950 0.954
[2,] 0.950 0.948 0.946 0.950 0.949 0.951
[3,] 0.951 0.952 0.948 0.949 0.951 0.953
[4,] 0.952 0.948 0.952 0.944 0.950 0.956

>     print(signif(fi.len, 3))
       [,1]  [,2]  [,3]  [,4]  [,5]  [,6]
[1,] 0.6170 3.080 7.420 20.70 43.00 88.30
[2,] 0.2680 1.180 2.760  7.51 15.40 31.60
[3,] 0.1360 0.586 1.350  3.66  7.51 15.30
[4,] 0.0552 0.234 0.535  1.44  2.96  6.02

}


\section{Discussion}
\label{S:discuss}

This paper considers the construction of an IM for exact prior-free probabilistic inference in a class of non-regular models.  The key to the development is the formulation of a certain differential equation whose solution identifies the feature of the underlying auxiliary variable to condition on.  In non-regular location parameter problems, say, not necessarily of the bounded-support type considered here, finding a solution to this differential equation that is free of the parameter is, at least in general, not an issue.  However, in other problems like the general parameter-dependent support considered in Section~\ref{S:nonregular}, one cannot generally expect that a solution is available that does not depend on the parameter.  In such cases, one can make use of the localization idea in \citet{imcond}.  This requires some sacrifice in the sense that the corresponding local (conditional) IM may not be valid for all possible assertions, but, in exchange, it does lead to IM-based plausibility intervals with exact coverage.  Our simulations also reveal that the exact IM intervals do not lose any efficiency, in terms of interval length, compared to the high-quality Bayes and fiducial intervals.  

Our focus here was on the case where the distribution had a bounded parameter-dependent support, in particular, the class of $\unif(a(\theta), a(\theta) + b(\theta))$ models.  However, the developments here would be applicable in other problems as well.  An interesting case is the non-regular regression model in \citet{smith1994}, where the error terms are iid from a distribution supported on the positive half-line, e.g., exponential.  We hope report elsewhere on some new IM developments in this and other non-regular examples.

\section*{Acknowledgments}

A portion of this work was completed while the first author was with the Department of Mathematics, Statistics, and Computer Science, University of Illinois at Chicago.

\bibliographystyle{apalike}
\bibliography{/Users/rgmartin/Dropbox/Research/mybib}

\begin{thebibliography}{}

\bibitem[Berger et~al., 2009]{bergerbernardosun2009}
Berger, J.~O., Bernardo, J.~M., and Sun, D. (2009).
\newblock The formal definition of reference priors.
\newblock {\em Ann. Statist.}, 37(2):905--938.

\bibitem[Hannig, 2009]{hannig2009}
Hannig, J. (2009).
\newblock On generalized fiducial inference.
\newblock {\em Statist. Sinica}, 19(2):491--544.

\bibitem[Hannig et~al., 2015]{hannig.review.2015}
Hannig, J., Iyer, H., Lai, R. C.~S., and Lee, T. C.~M. (2015).
\newblock Generalized fiducial inference: A review.
\newblock {\it J. Amer. Statist. Assoc.}, to appear.

\bibitem[Jin et~al., 2016]{jin.li.jin.2015}
Jin, H., Li, S., and Jin, Y. (2016).
\newblock The {IM}-based method for testing the non-inferiority of odds ratio
  in matched-pairs design.
\newblock {\em Statist. Probab. Lett.}, 109:145--151.

\bibitem[Martin, 2016]{gim}
Martin, R. (2016).
\newblock On an inferential model construction using generalized associations.
\newblock Unpublished manuscript, {\tt arXiv:1511.06733}.

\bibitem[Martin and Liu, 2013]{imbasics}
Martin, R. and Liu, C. (2013).
\newblock Inferential models: A framework for prior-free posterior
  probabilistic inference.
\newblock {\em J. Amer. Statist. Assoc.}, 108(501):301--313.

\bibitem[Martin and Liu, 2015a]{imcond}
Martin, R. and Liu, C. (2015a).
\newblock Conditional inferential models: Combining information for prior-free
  probabilistic inference.
\newblock {\em J. R. Stat. Soc. Ser. B}, 77(1):195--217.

\bibitem[Martin and Liu, 2015b]{imbook}
Martin, R. and Liu, C. (2015b).
\newblock {\em Inferential Models: Reasoning with Uncertainty}.
\newblock Monographs in Statistics and Applied Probability Series. Chapman \&
  Hall/CRC Press.

\bibitem[Shemyakin, 2014]{shemyakin2014}
Shemyakin, A. (2014).
\newblock Hellinger distance and non-informative priors.
\newblock {\em Bayesian Anal.}, 9(4):923--938.

\bibitem[Smith, 1994]{smith1994}
Smith, R.~L. (1994).
\newblock Nonregular regression.
\newblock {\em Biometrika}, 81(1):173--183.

\end{thebibliography}

\end{document}